\documentclass[prc,onecolumn,tightenlines,12pt]{revtex4}
\usepackage{graphicx}
\usepackage{dcolumn}
\usepackage{amsmath}
\begin{document}

\title{Deformations of the Tracy-Widom distribution}

\author{O. Bohigas$^{1}$, J. X. de Carvalho$^{2,3}$ and M. P. Pato$^{1,2}$}

\affiliation{$^{1}$CNRS, Universit\'e Paris-Sud, UMR8626, \\
LPTMS, Orsay Cedex, F-91405, France \\
$^{2}$Instituto de F\'{\i}sica, Universidade de S\~{a}o Paulo\\
Caixa Postal 66318, 05315-970 S\~{a}o Paulo, S.P., Brazil\\
$^{3}$Max-Planck-Institut f\"ur Physik komplexer Systeme\\
N\"othnitzer Stra$\beta$e 38, D-01187 Dresden, Germany}

\begin{abstract}

In random matrix theory (RMT), the Tracy-Widom (TW)
distribution describes  the behavior of the
largest eigenvalue. We consider here two models in which TW  
undergoes transformations.
In the first one disorder is introduced 
in the Gaussian ensembles by superimposing an external source 
of randomness. A competition 
between TW and a normal (Gaussian) distribution results, depending 
on the spreading of the disorder. 
The second model consists in removing at random a fraction of
(correlated) eigenvalues of a random matrix. The usual 
formalism of Fredholm determinants extends naturally. A continuous  
transition from TW to the Weilbull distribution, characteristc 
of extreme values of an uncorrelated sequence, is obtained.  
\end{abstract}
\maketitle

\section{Introduction}

In the beginning of the 90's, Tracy and Widom (TW) derived the 
probability distribution of the largest eigenvalue of random 
matrices belonging to the three Gaussian ensembles, the orthogonal 
(GOE), the unitary (GUE) and the symplectic (GSE)\cite{TW}. 
Few years later, Baik, Deift and Johansson proved that the 
longest increasing subsequence of a random permutation fluctuates 
as the largest GUE eigenvalue\cite{Baik} and triggered
TW applications in combinatorics and other areas
such as growing processes\cite{Johan,Grav,Rains,Spohn} 
(see \cite{TW1} for a review). 
The main ingredient in their derivation was the discovery 
that the formalism of random matrix theory based on Fredholm 
determinants and Painlev\'{e} equations\cite{Mehta} which
at the bulk of the spectrum is associated to integral 
equations with a sine-kernel, at the border of the 
spectrum it is associated to integral 
equations  with an Airy-kernel .
It is by now accepted that these distributions belong to 
universal class of extreme values of correlated sequences. 
Deviations from the TW have also been observed and studied. 
It has been found, for instance, that growing processes 
in random media may show regimes in which TW compete with other 
distributions\cite{Gravner}. 

In this general context it is important to establish links 
between TW and known universal distributions of 
extreme values of uncorrelated sequences.
In this last case, namely for a sequence of i.i.d. 
random variables, the probability that the extreme value 
is less than a given value $t$ is 
$\exp\left[-\int^{T}_{t}\rho(x)dx\right],$ 
where $\rho (x)$ is their density which extends
til $x=T$\cite{Coles}. Depending 
on the asymptotic behavior of the function $\rho(x),$ 
this probability distribution takes three forms. 
With $T=\infty,$ it becomes the Gumbel
distribution $\exp\left[-\exp(-y)\right]$ if $\rho$  
has a fast exponential decay and, if it decays 
with a power $\mu+1,$ it is the Fr\'{e}chet 
distribution $\exp(-1/y^{\mu})$ (in both cases $y$ 
is a properly scaled variable). With $T$ finite, 
that is if $\rho$ has a bounded support. it is the Weibull 
distribution $\exp(-y)$ where 
$y=\int^{T}_{t}\rho(x)dx$ has density one.
 
Largest eigenvalues of non-Gaussian ensembles 
(Wishart matrices) have been the subject of recent 
investigations\cite{Vivo}. 
It has been found that if the matrix elements are taken 
from a distribution with finite moments then the
TW holds\cite{Vivo}. Considering 
instead the case in which the distribution of the
matrix elements have long tails, it has been proven that 
when the second moment diverges, the largest 
eigenvalue and the largest matrix element follow a 
Fr\'{e}chet distribution with the 
same power $\mu\leq 2$\cite{Sosh}.
This result has more recently been extended til 
$\mu=4$\cite{Biroli}. 

Models to describe deviations from TW have been 
discussed by K. Johanson\cite{Johan1}. In one model,  
he studies the behavior of the largest eigenvalue 
of a matrix model\cite{Moshe} and finds that 
it is decribed by a kernel that goes from a Poisson 
kernel (see Eq. (\ref{3}) below) with an exponential 
density to the RMT Airy kernel. Accordingly, the 
distribution of the largest eigenvalue goes from 
Gumbel to TW. In another model, a deformed GUE ensemble 
is considered
in which the eigenvalue density fluctuates  
in such a way that the largest eigenvalue 
distribution goes from TW to Gaussian.  

Following similar lines to those of \cite{Johan1} 
the purpose of this note is to investigate other models 
that describe deformations of the TW. The first model results 
from superimposing to the Gaussian fluctuations an external 
source of randomness\cite{Josue}. This causes the eigenvalue 
semi-circle density to fluctuate and results in features  
common to growing processes in random media. 
The second model is based on the recent recognition 
that the mathematical structure of random matrix 
theory (RMT) also describes the statistical properties 
of the eigenvalues of spectra when a fraction of eigenvalues 
is randomly removed \cite{Boh1}. As this 
operation reduces correlations, it describes intermediate 
statistics between RMT and Poisson statistics. 
Focusing on eigenvalues at the edge of the spectrum we show
here that this leads to a transition 
from TW to a Weibull distribution. 

Consider the Gaussian random matrix ensembles defined by a
density distribution 

\begin{equation}
P_{G} (H;\alpha) =\left(\frac{\alpha\beta}{\pi}\right)^{f/2}
\exp(-\alpha\beta\mbox{tr} H^2) \label{717}
\end{equation}
where  $f=N+\beta N(N-1)/2$ is the number of independent 
matrix elements and $\beta $ is the Dyson index that takes the
values $1,2,$ and $4$ for the Orthogonal (GOE), the Unitary (GUE) and
the Symplectic (GSE) ensembles, respectively. In (\ref{717}), 
the normalization constant is calculated with respect to the measure
$dH=\prod_{1}^{N}dH_{ii}\prod_{j>i}\prod_{k=1}^{\beta}
\sqrt{2}dH^{k}_{ij}.$

Let us start by recalling known facts about the eigenvalues of these 
Gaussian ensembles including some recent results regarding the behavior 
of their largest values. It is well known, for instance, that, 
to leading order, their eigenvalue density is given by the 
Wigner's semi-circle law

\begin{equation}
\rho(\lambda)=\left\{ 
\begin{array}{rl}
\frac{1}{2\pi\sigma^2} \sqrt{4N\sigma^2-\lambda^2} , & 
\mid\lambda\mid<2\sigma\sqrt{N} \\ 
0, & \mid \lambda \mid >2\sigma\sqrt{N}  \\ 
\end{array}
\right. \label{9a}
\end{equation}  
where $\sigma=1/\sqrt{4\alpha}$ is the variance of the off-diagonal matrix
elements. To study the behavior of the largest eigenvalues in the limit 
of large matrix size $N,$ one introduces the scaling

\begin{equation}
\lambda =(2\sqrt{N}+\frac{s}{N^{1/6}})\sigma \label{817}
\end{equation}
that substituted in (\ref{9a}) leads to the $N$-independent density

\begin{equation}
\rho(s)=\left\{ 
\begin{array}{rl}
\frac{1}{\pi}\sqrt{-s}, & s\leq 0 \\ 
0, & s>0  \\ 
\end{array}
\right. \label{9}
\end{equation}
at the edge of the spectrum. In the scaled variable $s,$ 
the probabilities $E\left(k,s\right)$ 
with $k=0,1,2,..$ that the infinite interval $(s,\infty)$ has, 
respectively $k$ eigenvalues, are obtained from the generating 
function $G(s,z)$ through the relation \cite{TW,For}
\begin{equation}
G(s,z)=\sum_{n=0}^{\infty }\left(-1\right)^{n}\left(z-1\right)^{n}
E\left(n,s\right) \label{585}
\end{equation}
such that 

\begin{equation}
E(n,t)=\frac{\left(-1\right)^{n}}{n!}\left[\frac{\partial G(s,z)}
{\partial z^{n}}\right]_{z=1}.
\end{equation}
For the three symmetry classes, the generating functions
$G_{\beta}(s,z)$ with $\beta =1,2$ and $4$ have been derived.
Starting with the unitary case, $G_{2}(s,z)$ can be identified with 
the Fredholm determinant associated to the integral operator acting on
the interval $(s,\infty)$ with kernel\cite{Forrester} 

\begin{equation}
K(x,y)=\frac{\mbox{Ai}(x)\mbox{Ai}^{\prime}(y)-{\mbox{Ai}(y)
\mbox{Ai}^{\prime}(x)}}{x-y} \label{818}
\end{equation}
where $ \mbox{Ai}(s)$ is the Airy function. $G_{2}(s,z)$ is given by

\begin{equation}
G_{2}(s,z) =\exp\left[-\int_{s}^{\infty}
(x-s)q^{2}(x,z)dx\right]
\label{148}
\end{equation}
where $q(s,z)$ satisfies the  Painlev\'{e} II equation

\begin{equation}
q^{\prime\prime} = s q +2q^3
\label{149}
\end{equation}
with boundary condition 

\begin{equation}
q(s,z)\sim \sqrt{z}\mbox{Ai}(s) \mbox{  when  } s 
\rightarrow \infty. \label{418}
\end{equation} 
For GOE ($\beta=1$) and GSE ($\beta=4$) the generating functions
are\cite{Dieng}

\begin{equation}
\left[G_{1}(s,z)\right]^2 =G_{2} (s,\bar{z})
\frac{z-1-\cosh\mu (s,\bar{z})+\sqrt{\bar{z}}
\sinh\mu (s,\bar{z})}{z-2}
\end{equation}
and

\begin{equation}
\left[G_{4}(s,z)\right]^2 =G_{2} (s,z)
\cosh^{2} \frac{\mu (s,z)}{2}
\end{equation}
where $\bar{z} = 2z-z^{2}$ and

\begin{equation}
\mu (s,z) =\int_{s}^{\infty} q(x,z)dx. \label{444}
\end{equation}
We remark that the above expression for the GSE case was obtained
in Refs. \cite{Forrester,Dieng} using a scaling that assume 
$N/2$ eigenvalues.

The above equations give a complete description of the fluctuations 
of the eigenvalues at the edge of the spectra of the Gaussian
ensembles. In particular, for the largest eigenvalue, the TW 
distributions are expressed in terms of these generating functions
as $E_{G,\beta}\left(E_{max} < \lambda\right)=G_{\beta}(s,1). $

\section{Disordered ensembles}

To investigate the modifications these probabilities undergo when
an external source of randomness is superimposed to the Gaussian 
fluctuations, we consider disordered ensembles whose matrices,  
$H(\xi,\alpha),$ are defined as\cite{Josue}

\begin{equation}
H(\xi,\alpha) =\frac{ H_G(\alpha)}{\sqrt{\xi/{\bar \xi}}}
\label{1}
\end{equation}
where $H_G$ is a matrix of (\ref{717})
and $\xi$ is a positive random variable with distribution   
$w(\xi )$ with average ${\bar \xi}$ and variance $\sigma_{w}.$  
From (\ref{1}) and (\ref{717}) it is deduced that the joint density 
distribution of the matrix elements is a superposition of the 
Gaussian ensembles distributions weighted with $w(\xi ),$ namely

\begin{equation}
P\left( H;\alpha\right) =
\int d\xi w( \xi ) 
P_{G}(H;\alpha\xi/{\bar \xi}).  \label{400}
\end{equation}

Changing variables from matrix elements to eigenvalues and
eigenvectors, it is also found, after integrating out the
eigenvectors, that the joint probability distribution of the 
eigenvalues is obtained by averaging over the joint distribution of
the Gaussian ensembles. As a consequence, measures of this average
ensemble are averages over the Gaussian measures. The eigenvalue
density, for instance, turns out to be an average over
Wigner's semi-circles with different radii, that is

\begin{equation}
\rho \left(\lambda;\alpha \right) = \int d\xi w(\xi)
\sqrt{4\sigma^2(\xi)  N-\lambda ^{2}}/\left[2\pi\sigma^2(\xi)
\right],\label{126}
\end{equation}
where the $\xi$-dependent variance, $\sigma(\xi)$ is given by
\begin{equation}
\sigma(\xi)=\sigma\sqrt{\bar{\xi}/\xi} \label{446}
\end{equation}

The probability that the largest eigenvalue $\lambda_{max}$ is 
smaller than  a given value $t$ can be calculated by evaluating 
the probability that the interval $(t,\infty)$ is empty. This is
obtained by integrating the joint probability distribution of the
eigenvalues in the interval $(-\infty,t)$ over all eigenvalues,
we find

\begin{equation}
E_{\beta}\left(\lambda_{max} < t\right)=
\int d\xi w (\xi )
E_{G,\beta}\left[S(\xi,t)\right] 
\label{146}
\end{equation}
with the argument of $S(\xi,t)$ obtained by plugging in 
Eq. (\ref{817}) the above $\xi$-variance, Eq. (\ref{446}), 
namely

\begin{equation}
S(\xi,t)=N^{1/6}\left[\frac{t}{\sigma(\xi)}-2\sqrt{N}\right].
\label{147}
\end{equation}

Equations (\ref{146}) and (\ref{147})  
give a complete analytical description of the
behavior of the largest eigenvalue once the function $w(\xi)$ is
chosen. But even without specifying  $w(\xi),$ asymptotic results 
can be derived by comparing its localization, given by the 
ratio $\sigma_{w}/\bar{\xi},$ with that of $E_{G,\beta}$ considered as
functions of the integrand variable $\xi$ through Eq. (\ref{147}). 
Since the widths of these last ones depend on the matrix size $N,$
let us introduce a positive parameter $z$ such that  
$\sigma_w /\bar{\xi}= N^{-z}$ which is kept fixed when 
the limit $N\rightarrow \infty$ is taken. 

As $z>0,$ when $N$ increases, the $w(\xi)$ distribution becomes more 
and more localized and, if asymptotically it can be approximated by
a Gaussian, by changing the integration variable to 

\begin{equation}
\xi =\bar{\xi} - v\sigma_{w}, \label{667}
\end{equation}
Eq. (\ref{146}) can be written as 

\begin{equation}
E_{\beta}\left(\lambda_{max} < t\right)=\frac{1}{\sqrt{2\pi}} 
\int_{-\infty}^{\infty}dv \exp \left( -\frac{v^2}{2}\right)
E_{G,\beta}\left[S(v,t)\right] , \label{666}
\end{equation}
where the argument $S(v,t),$ after neglecting higher order terms in $1/N,$ 
takes the form

\begin{equation}
S(v,t) =N^{1/6}\left(\frac{t}{\sigma}- v N^{1/2-z}-2\sqrt{N}\right)=
s- N^{2/3-z}v. 
\label{636}
\end{equation}
In the last step of (\ref{636}), Eq. (\ref{817}) was used and 
it was assumed that $z>2/3.$ Taking now 
the limit $N\rightarrow \infty$, the vanishing of the second term in the 
r.h.s. of (\ref{636}) makes 
$S(v,t)$ independent of $v$ and the distributions $E_{G,\beta}(s),$ 
i.e. TW, are recovered. 
Values of the parameter $z$ greater than $2/3$ correspond to situations 
in which the distribution $w(\xi)$ collapses faster than $E_{G,\beta},$ 
for $z<2/3,$ on the other hand, it is the opposite
that happens. To be able to get $N$-independent results in this range of
values of $z,$ it is necessary to modify the scaling to 

\begin{equation}
s = N^{z-1/2}\left(\frac{t}{\sigma}-2\sqrt{N}\right),
\end{equation}
in which case, the argument $S\left[v,t(s)\right]$ becomes 

\begin{equation}
S\left[v,t(s)\right] =N^{2/3-z}(s-v).
\end{equation}
Taking now the limit of $N\rightarrow \infty$,   
$E_{G,\beta}(S) $ becomes a step function centered at $v=s$ and
the distribution goes to the normal distribution $N(0,1).$ 
Finally, at the critical value $z=2/3,$ from both sides, (\ref{666}) converges
to the convolution of the normal distribution and TW. In summary we have the
three regimes 

\begin{equation}
E_{\beta}\left(\lambda_{max} < t\right)=\left\{ 
\begin{array}{rl}
E_{G,\beta}(s)    , & z > 2/3 \\ 
\frac{1}{\sqrt{2\pi}} 
\int_{-\infty}^{\infty}dv \exp \left( -\frac{v^2}{2}\right)
E_{G,\beta}(s-v) , & z=2/3  \\
\frac{1}{\sqrt{2\pi}} 
\int_{-\infty}^{s}dv \exp \left( -\frac{v^2}{2}\right) , & z < 2/3  \\       
\end{array}
\right. 
\end{equation}  

Consider now the case in which the distribution $w(\xi)$ 
is independent of $N.$ 
In this case, it is appropriate to make the parameter $\alpha$ 
equal to the matrix size $N.$ With this scaling, 
the eigenvalues of the matrices of the average ensemble are
located in the interval $(-1,1),$ and the $\xi$-dependent argument
$S(\xi,t),$ Eq. (\ref{147}), takes the simple form    
$S(\xi,t)= 2N^{2/3}(t\sqrt{\xi/\bar{\xi}}-1).$ This expression 
makes evident that when the matrix size $N$ increases,
for the three invariant ensembles, 
the function $E_{G,\beta}$ become a step function 
centered at $\xi=\bar{\xi}/t^2.$ Therefore, in this regime, 
the probability distribution for the largest eigenvalue converges to

\begin{equation}
E_{\beta}\left(\lambda_{max} < t\right)=\int_{\bar{\xi}/t^2}^{\infty}d\xi 
w\left(\xi \right)  
\label{164}
\end{equation}
with density

\begin{equation}
\frac{dE_{\beta}\left( t\right)}{dt} =
\frac{2 \bar{\xi}w(\bar{\xi}/t^2)}
{ t^{3}} . \label{605}
\end{equation}
A special choice of $w(\xi)$ that already has appeared in previous studies 
of disordered ensembles\cite{Bertuola,Gernot}, is that in which it is the one
parameter family of Gamma distributions 

\begin{equation}
w(\xi)=
\exp(-\xi) \xi^{\bar{\xi} -1} /\Gamma(\bar{\xi}).  \label{18}
\end{equation} 
In this case, (\ref{605}) becomes

\begin{equation}
\frac{dE_{\beta}\left( t\right)}{dt} =
\frac{2 \bar{\xi}^{\bar{\xi}}\exp(-\bar{\xi}/t^2)}
{\Gamma(\bar{\xi}) t^{2\bar{\xi}+1}} . \label{603}
\end{equation}
that defines a long-tailed distribution of extreme values of a 
correlated set of points(we remark that this distribution distribution 
has recently been considered in random covariant matrices\cite{Potters}).

Notice that in (\ref{605}) the variable $t$ is the eigenvalue
itself without any edge scaling. This means that, in this case, 
fluctuations become of the 
order of the size of the average ensemble spectrum. 
Although (\ref{605}) resembles a  Fr\'{e}chet distribution, the 
fact that the power of $t$ in the exponent is fixed at the 
value two makes it a different distribution. 
Nevertheless, for $\bar{\xi}=1$ 
it is indeed a Fr\'{e}chet distribution. We remark that 
$\bar{\xi}=1$ corresponds to the critical distribution of 
the family defined by Eq. (\ref{18}) that separates the 
ones that converge from those that diverge at the origin. The  
asymptotical power-law decay of (\ref{603}), similar to that of 
Fr\'{e}chet distribution, suggest that in the asymptotic region 
the extreme value behaves independently of the other eigenvalues
while, in the internal region, the presence of the others 
are felt.

\section{From Tracy-Widom to Weibull}

Let us now turn to the case of a model to describe largest eigenvalues
of spectra in the intermediate regime between RMT and Poisson. This
model is brought on by the fact that the generating function, 
Eq. (\ref{585}), can be interpreted as a probability. Indeed, assuming
with $0<z<1$ that the factor $(1-z)^n$ that multiplies $E(n,s)$ 
in (\ref{585}) is the
probability that the $n$ eigenvalues in the interval $(s,\infty)$ have
been removed, then summing all the terms gives the probability 
that there is no level in the interval.  

A realization of this situation was considered in
Ref. \cite{Boh} in which the effect of removing at random 
a fraction $1-f$ of eigenvalues of RMT spectra was investigated. In
this case, $1-f$ is the probability that a given eigenvalue has been 
dropped from the spectrum. Therefore with the identification of $z$ with $f$
the generating function  Eq. (\ref{585}) becomes the probability
distribution of the largest eigenvalue for this kind of randomly 
incomplete spectra.   

In Refs. \cite{Boh,Boh1}, in studying the effect of 
incompleteness in the spectral statistics at the bulk, an interval of
length $s$ is increased by a factor of $1/f$ to compensate the
reduction in the average number of levels inside it. 
Following the same idea at the edge, we want a scaling of the variable $s$ such
that the average number of
eigenvalues in the interval $(s,\infty)$ remains the same when a fraction of
levels is removed. This average is 

\begin{equation}
<n>=\frac{2}{3\pi}\left(-s\right) ^{3/2} , \label{35}
\end{equation}  
obtained by integrating (\ref{9})  from $s$ to $\infty.$ 
Therefore, in order to keep it invariant when the density of eigenvalues
is reduced by a factor of $f,$ $s$ has to be divided by $f^{2/3}.$

Using this scaling in the Airy-kernel, Eq. (\ref{818}), we expect that
when the limit $f\rightarrow 0$ is taken  
it converges to the Poisson-kernel \cite{Johan}

\begin{equation}
K_P (x,y)= \left\{ 
\begin{array}{rl}
0, & x\neq y \\ 
\rho(x), & x = y \\ 
\end{array}
\right. \label{3}
\end{equation}
with density given by (\ref{9}). In fact, we have

\begin{equation}
\lim _{f\rightarrow 0}
\frac{\mbox{Ai}(x/f^{2/3})\mbox{Ai}^{\prime}(y/f^{2/3})
-{\mbox{Ai}(y/f^{2/3})
\mbox{Ai}^{\prime}(x/f^{2/3})}}{(x-y)/f^{2/3}}=K_P (x,y).
\end{equation}

Denoting by  $\hat{e}(s,f)$ the probability that the semi-infinite 
interval $(s,\infty)$ is empty, for the incomplete spectra it is given
by

\begin{equation}
\hat{e}_\beta (s,f)=\sum_{k=0}^{\infty }
\left(1-f\right)^{k} E_\beta (k,s/f^{2/3})= G_\beta (s/f^{2/3},f), \label{15}
\end{equation}
that means that in an incomplete spectrum the largest eigenvalue can
be anyone other of the $n$th largest eigenvalues. The last equality 
in  (\ref{15}) follows from Eq. (\ref{585}) and shows that the 
generating functions for the three symmetry classes contain
a comprehensive description of the largest eigenvalues of complete 
and incomplete RMT spectra.

To investigate the limit $f\rightarrow 0$ we remark that as the scaling 
factor $f^{2/3}$ appears in the denominator, 
for small $f$ the function  $q(x,z)$ can be replaced by its asymptotic 
form at $x\rightarrow \pm\infty$. For $x>0,$ this is given by the 
exponential decay of the asymptotic behavior of the Airy function. 
So, at this positive side, $q(x,z)$ vanishes and 
$\hat{e}_{\beta}(s,f)=1$. For $s<0,$ by the same argument,
the integrals can be performed from $s$ to zero with $q(x,z)$ 
replaced by its asymptotic expression for large negative values. 
For $0<z<1,$ this
expression has been worked out by 
Hastings and McLeod\cite{Hastings}, that found 

\begin{equation}
q(x,z) \sim d (-x)^{-1/4} \sin \left[ \frac{2}{3}(-x)^{3/2}-
\frac{3}{4}d^{2}\log(-x)-c\right]\label{40}
\end{equation}
where $d^{2} =-\frac{1}{\pi}\log(1-z).$
 
The integral in (\ref{148}) contains the square of $q(x,z)$ so, if we use
the relation $2\sin^{2}x=1+\cos2x,$ we find that the generating 
functions have a smooth and an oscillating part. 
As the period and the amplitude of the oscillations decrease with $f,$ 
in the limit of small $f,$ the oscillating part averages out
to zero and can be neglected. For the same reason, in this limit, 
the function $\mu (s,f),$ Eq. (\ref{444}), vanishes.
Taking then all this into account and after evaluating integrals, we 
find that when $f\rightarrow 0,$  the largest eigenvalue
distributions, for the three symmetry classes, converge 
to the Weibull distribution 

\begin{equation}
\hat{e_\beta }(s)=\left\{ 
\begin{array}{rl}
\exp\left[-\frac {2g_\beta}{3\pi}(-s)^{3/2}\right], & s\leq 0 \\ 
1, & s > 0 \\ 
\end{array}
\right. , \label{44}
\end{equation}
where $ g_\beta =1 $ for $\beta=1,2$ and   
$ g_\beta=1/2 $ for $\beta=4$ (this parameter reflects the 
fact mentioned above that, in the GSE case, we are using a 
scaling with $N/2$).
This Weibull distribution describes the extreme value of a set 
of uncorrelated points with a semi-circle density distribution. 
In Fig. 1, the transition from TW  ($f=1$) to Weibull ($f=0$) 
is illustrated for the GUE case ($\beta=2$). 
The oscillations at intermediate values are clearly seen.

\section{Conclusion}
 
In conclusion, we have investigated the behavior of the largest
eigenvalue of two models that generalize the RMT Gaussian
ensembles. In the first one, by introducing disorder in the ensemble,
the eigenvalue density is lead to fluctuate generating at the edge a 
competition between the Tracy-Widom distribution and the
normal distribution. This kind of behavior has been observed in
growing processes in random media. In the second model, the 
eigenvalue density is kept fixed but the correlations 
among the eigenvalues are progressively reduced. It is then 
observed a continuous transition from the TW to the Weibull 
distribution, characteristic of uncorrelated variables.  

This work is supported in part by the Brazilian agencies CNPq and FAPESP.

{\bf Figure Captions}

Fig. 1  Density distribution of largest eigenvalue for GUE
($\beta=2$). For complete sequence ($f=1,$ Tracy-Widom), for 
uncorrelated sequence (Weibull, limit $f\rightarrow 0$;
Eq. (\ref{44})) and partially incomplete sequences 
( $f=0.3, 0.6$, from  (\ref{15})).

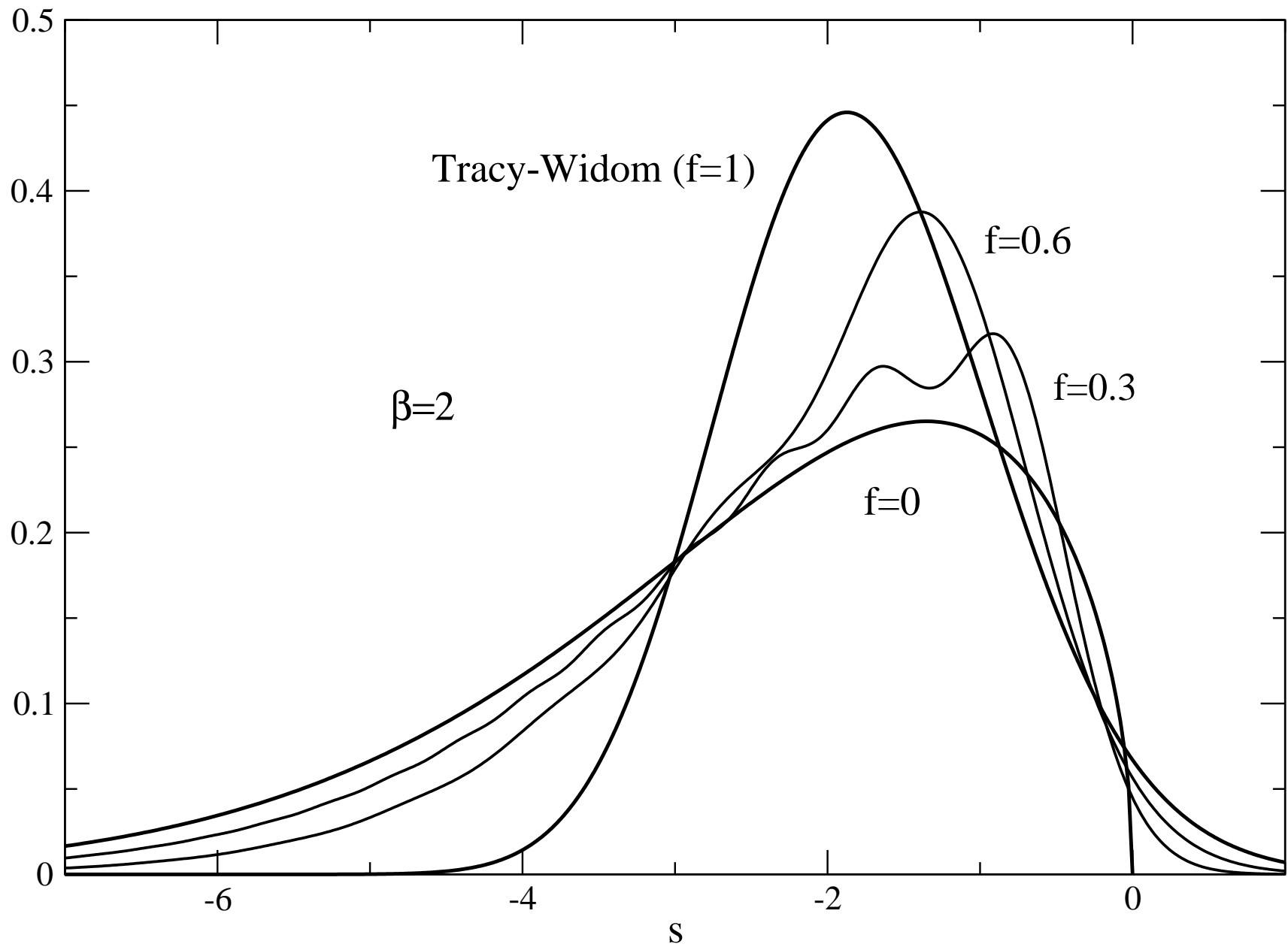
\end{document}